\documentclass [letterpaper,11pt]{article}

\usepackage{amsmath, amsthm}
\usepackage{amssymb}
\usepackage{graphicx}
\usepackage{verbatim}

\begin{document}

\title{\centering\Large{
On Revenue Monotonicity in Combinatorial Auctions
                               }}
\date{}

\author{Andrew Chi-Chih Yao\thanks{Institute for Interdisciplinary Information Sciences, Tsinghua University, Beijing.}}
\maketitle{}

\begin{abstract}
Along with substantial progress made recently in designing near-optimal mechanisms for multi-item auctions, interesting structural questions have also been raised and studied. In particular, is it true that the seller can always extract more revenue from a market where the buyers value the items higher than another market?  In this paper we obtain such a revenue monotonicity result in a general setting. Precisely, consider the revenue-maximizing combinatorial auction for $m$ items and $n$ buyers in the Bayesian setting, specified by a valuation function $v$ and a set $F$ of $nm$ independent item-type distributions. Let $REV(v, F)$ denote the maximum revenue achievable under $F$ by any incentive compatible mechanism.  Intuitively, one would expect that $REV(v, G)\geq REV(v, F)$ if distribution $G$ stochastically dominates $F$.  Surprisingly, Hart and Reny (2012) showed that this is not always true even for the simple case when $v$ is additive.  A natural question arises: Are these deviations contained within bounds?  To what extent may the monotonicity intuition still be valid?  We present an {approximate monotonicity} theorem for the class of fractionally subadditive (XOS) valuation functions $v$, showing that $REV(v, G)\geq c\,REV(v, F)$ if $G$ stochastically dominates $F$ under $v$ where $c>0$ is a universal constant. Previously, approximate monotonicity was known only for the case $n=1$: Babaioff et al. (2014) for the class of additive valuations, and Rubinstein and Weinberg (2015) for all subaddtive valuation functions.
\end{abstract}

\newpage

\section{Introduction}

Along with substantial progress made recently in designing near-optimal mechanisms for multi-item auctions, interesting structural questions have also been raised and studied. In particular, is it true that the seller can always extract more revenue from a market where the buyers value the items higher than another market?  In this paper we obtain such a revenue monotonicity result in a general setting, leveraging on recent progress made in the mechanism design literature.

In the simplest case of Myerson's 1-item auction \cite{Myerson1981}, let $REV(\mathcal{F})$ denote the optimal revenue for independent valuation distributions $\mathcal{F}=F_1\times \cdots \times F_n$.  Is it true that $REV(\mathcal{G})\geq REV(\mathcal{F})$ when $\mathcal{G}=G_1\times \cdots \times G_n$ stochastically dominates $\mathcal{F}$ (i.e. $G_i$ stochastically dominates $F_i$ for each buyer $i$)?  Intuitively, if each buyer $i$ is prepared to pay more for the item, it seems reasonable that the seller should be able to extract more revenue.  This is indeed true for the 1-item auctions, as remarked in Rubinstein and Weinberg \cite{RW}, as a consequence of Myerson's characterization.

The revenue monotonicity question becomes much subtler when there are $m >1 $ items in the auction.  Hart and Reny \cite{HR2015} showed that revenue monotonicity is not universally true even with just one buyer ($n=1$) and two items ($m=2$). They gave examples with distributions $\mathcal{G}$ stochastically dominating $\mathcal{F}$, yet $REV(\mathcal{G})< REV(\mathcal{F})$.  Thus, when there are $m>1$ items, the target can only be \emph{approximate revenue monotonicity}, e.g. $REV(\mathcal{G})\geq c REV(\mathcal{F})$ for some absolute positive constant $c$.

The following monotonicity results have been shown for $n=1$ buyer and any number of items:  if $\mathcal{G}$ stochastically dominates $\mathcal{F}$, then $REV(\mathcal{G})\geq \frac{1}{6} REV(\mathcal{F})$ when the valuation function is additive (Babaioff et al. \cite{Baba2014}), and $REV(\mathcal{G})\geq \frac{1}{338} REV(\mathcal{F})$ for combinatorial auctions with any subadditive valuations (Rubinstein and Weinberg \cite{RW}). These results were obtained as immediate corollaries of their respective near-optimal mechanisms which are revenue monotone in distributions. Recently, near-optimal mechanism were found for any $n,m$ in Yao \cite{Yao2015} and Cai, Devanur and Weinberg \cite{CDWSTOC2016} for additive valuations, and in Cai and Zhao \cite{CZ2017} (also Feldman, Gravin and Lucier \cite{FGL2015} for welfare maximization) for XOS valuation functions. However, these mechanisms are not obviously revenue monotone in distributions; as such, no general monotonicity results are known for $n>1$.

Our main result is to resolve this question in the affirmative for the class of XOS valuation functions. For any $m, n$, and combinatorial auctions with XOS valuations $v$, we show that $REV(\mathcal{G})\geq c REV(\mathcal{F})$ for $c=\frac{1}{1448}$ if $\mathcal{G}$ stochastically dominates $\mathcal{F}$ with respect to $v$. We also prove two auxiliary theorems which are needed in proving our main result, and also useful in their own right. Firstly, for any single-parameter environment auction $A\subseteq [0,1]^n$, we show that the optimal revenue satisfies $REV_A(\mathcal{G})\geq  REV_A(\mathcal{F})$ if $\mathcal{G}$ stochastically dominates $\mathcal{F}$.  This implies that revenue monotonicity is true not just for Myerson's 1-item optimal auctions, but also for general 1-dimensional auction problems in which the allocation vectors are restricted to an arbitrary allowable set of patterns.  Secondly, as a consequence of the single-parameter monotonicity above, we infer that $REV(\mathcal{G})\geq \frac{1}{24} REV(\mathcal{F})$ in the unit-demand multi-item auctions.

\noindent{\textbf{Contributions of this work:}} \ 1) We have given answers of great generality to the revenue monotonicity question, applicable to all XOS subadditive valuation functions, while previously little was known even for the additive valuations.  2) Our key innovation in proof technique is a conceptual one (see Section 4.2), requiring no complicated calculations or analysis. A main difficulty in proving XOS monotonicity is the fact that the near-optimal mechanism given in \cite{CZ2017} is not monotone. To overcome this obstacle, we first embed our auction into a more relaxed context (that of \emph{digital goods}).  In this larger space, we can then establish revenue monotonicity via a connecting path (in the new space) between the two embedded distributions. 3) At a more philosophical level, we agree with the sentiment (such as expressed in Hart and Nisan \cite{HN2012}) that the goal of designing mechanisms is not only to produce an algorithm that works, but also to reveal mathematical structures that allow interesting questions such as revenue monotonicity to be answered. Our present work serves as another validation of the fruitfulness of this approach.

\section{Main Results}

We present results in three standard auction models in the {\it independent setting}, in which all the $mn$ item-types are drawn from independent distributions. Some familiar terminologies are reviewed below.

For any two random variables $X, Y$,  we write $X\simeq Y$ if $X$ and $Y$ are \emph{equal in distribution}, i.e., $Pr\{X\in S\}=Pr\{Y\in S\}$ for any measurable set $S$. Let $F$, $G$ be distributions over $[0, \infty)$.  We write $F\preceq G$ if ${F}$ is \emph{(stochastically) dominated} by ${G}$ (i.e. $Pr\{F>t\}\leq Pr\{G>t\}$ for all $t\in [0, \infty)$). Equivalently, we may write $G\succeq F$ if $G$ dominates $F$. Let $\mathcal{F}=F_1\times \cdots \times F_n$ and $\mathcal{G}=G_1\times \cdots \times G_n$ be product distributions over $[0, \infty)^n$. We say $\mathcal{F}\preceq \mathcal{G}$ (or equivalently $\mathcal{G}\succeq \mathcal{F}$) if $F_i \preceq G_i$ for each $i$.

Let $M=(x,p)$ be an $n$-buyer DIC-IR mechanism, and let $\mathcal{F}$ be an input valuation distribution over $[0, \infty)^n$.  Let $X_{M, \,\mathcal{F}}$ and $P_{M,\, \mathcal{F}}$ stand for the random variables $X_{M, \,\mathcal{F}}=x(t)$,  $P_{M, \,\mathcal{F}}=p(t)$ in the probability space $\{t|\, t\sim \mathcal{F}\}$.

A \emph{randomized} DIC-IR mechanism is a family of mechanisms $\{M^r|\, r\sim \mathcal{H}\}$, where each $M^r = (x^r, p^r)$ is a DIC-IR mechanism, and $r$ is randomly chosen according to some distribution $\mathcal{H}$.  Let $x^r=(x^r_1, \ldots, x^r_n)$, $p^r=(p^r_1, \ldots, p^r_n)$.  The \emph{revenue} yielded by $M^R$ under input distribution $\mathcal{F}$ is defined as $M^R(\mathcal{F})=\sum^n_{i=1}E_{r\sim \mathcal{H},\, t\sim \mathcal{F}}(p^r_i(t))$.  Let $X_{ M^R,\,\mathcal{F}}$ and $P_{ M^R,\,\mathcal{F}}$ stand for the random variables $X_{ M^R,\,\mathcal{F}}=x^r(t)$, $P_{ M^R,\,\mathcal{F}}=p^r(t)$ in the probability space
$\{(r,t) |\, r\sim \mathcal{H}, t\sim \mathcal{F}\}$.

A \emph{single-parameter environment} (see e.g. Gonczarowski and Nisan \cite{GN2017})  is specified by a set of \emph{possible outcomes} $A\subseteq [0,1]^n$. For a valuation distribution $\mathcal{F}$ over $[0, \infty)^n$, $REV_A(\mathcal{F})$ is defined as the maximum revenue yielded by any DIC-IR mechanism, where the allocation $x(t)$ for any type profile $t$ is restricted to be in the set $A$.

\vskip 8pt

\noindent{\textbf{Theorem 1}}  [{Single-Parameter Environment}]  Let $M=(x, p)$ with allocation function $x$ and payment function $p$ be an $n$-player DIC-IR mechanism with valuation distribution  $\mathcal{F}$ over $[0, \infty)^n$. Let $X_{M, \,\mathcal{F}}, P_{M,\,\mathcal{F}}$ denote the random variables corresponding to $x, p$. Then for any valuation distribution  $\mathcal{G}\succeq \mathcal{F}$, there exists a randomized DIC-IR mechanism $M^R$ such that  \\
(A) $(X_{ M^R,\,\mathcal{G}}, P_{ M^R,\,\mathcal{G}})\simeq \{(x(t), p_G(t))\,|\, t\sim \mathcal{F}\}$ where $p_G: [0, \infty)^n \rightarrow [0, \infty)^n$ and $p_G\geq p$.\\
(B) $X_{ M^R,\,\mathcal{G}}\simeq X_{M,\,\mathcal{F}}$, and $M^R(\mathcal{G})\geq M(\mathcal{F})$.

\vskip 8pt

\noindent{\textbf{Corollary}}  For any single-parameter environment $A$, we have $REV_A(\mathcal{G})\geq REV_A(\mathcal{F})$ if
$\mathcal{G}\succeq \mathcal{F}$.
\vskip 8pt

Theorem 1 can be used to prove a monotonicity theorem for unit-demand multi-item auctions. Let $DREV^{UD}(\mathcal{F})$ denote the optimal revenue achievable by any deterministic DIC-IR mechanism for distribution $\mathcal{F}$ in the unit-demand model, and let  $REV^{UD}(\mathcal{F})$ be the optimal revenue achievable by any incentive-compatible mechanism allowing randomized lotteries. It is known (Chawla, Malec and Sivan \cite{CMS2015}) that allowing lotteries sometimes can generate more revenues.
\vskip 8pt

\noindent{\textbf{Theorem 2}\ [Unit-Demand Multi-item Auction]  If $\mathcal{G}\succeq \mathcal{F}$, then
\begin{align*}
DREV^{UD}(\mathcal{G})\geq \max\{\frac{1}{6} DREV^{UD}(\mathcal{F}), \frac{1}{24} REV^{UD}(\mathcal{F})\}.
\end{align*}

\vskip 8pt

Theorem 2 is useful in the proof of the next theorem, which is the main result of this paper. Let $v=(v_1, \cdots, v_n)$ be a valuation function, and each $v_i$ has $\alpha_v (\geq 1)$-{\it supporting prices} for all $t_i$ (see Section 5 for definitions). It is known that $\alpha_v =1$ if $v$ is fractionally subadditive (XOS), and more generally $\alpha_v = O(\log m )$ if $v$ is subadditive. Let $D$ and $D'$ be distributions over some type space. We say that $D$ {\it v-stochastically dominates} $D'$ if a coupled random pair $(t, t')$ can be generated such that the marginal distributions satisfy $t \sim D$, $t' \sim D'$ and $v(t, S)\geq v(t', S)$ for all $S\subseteq [m]$.

In the next theorem, $REV_{DIC}(v, \mathcal{G})$ refers to the maximum revenue achievable by any deterministic DIC-IR mechanism under valuation $v$ and distribution $\mathcal{G}$; $REV_{BIC}(v, \mathcal{F})$ refers to the maximum revenue achievable by any randomized BIC-BIR mechanism under valuation $v$ and distribution $\mathcal{F}$.

\vskip 8pt
\noindent{\textbf{Theorem 3}\ [Subadditive Combinatorial Auction] Let $v$ be a valuation satisfying monotonicity, subadditivity and with no externalities.  If $\mathcal{G}\succeq_v \mathcal{F}$, then for any $0<b<1$ we have
 \begin{align*}REV_{DIC}(v, \mathcal{G})\geq \frac{1}{\lambda} REV_{BIC}(v, \mathcal{F})
 \end{align*}
 where $\lambda= 32 \alpha_v+6\left(12+  \frac{8}{1-b}+\alpha_v(\frac{16}{b(1-b)} + \frac {96}{1-b})\right)$.

\vskip 8pt
\noindent{\textbf{Corollary}} \ If $v$ is XOS, then $\alpha_v=1$ and choosing $b=\frac{1}{4}$ gives
  \begin{align*}REV_{DIC}(v, \mathcal{G})\geq \frac{1}{1448}\ REV_{BIC}(v, \mathcal{F}).
 \end{align*}

Theorem 1 is easy to prove for the special case when all the distributions $F_i, G_i$ over $[0,\infty )$ are continuous and strictly increasing, using Myerson's theory. The general case needs greater care, due to discontinuities and plateaus of the distributions. We omit the proof here.

We prove Theorems 2 and 3 in Sections 3 and 4, respectively. More background information and discussions will be presented along the way in proving these results.

\section{Unit-Demand Auctions}

Consider revenue maximization in an auction with $m$ heterogenous items to sell, and $n$ buyers who are \emph{unit-demand}, i.e., each buyer is allocated either $0$ or $1$ item.  Buyer $i$ has for item $j$ valuation distribution $F_{ij}$ over $[0,\infty)$, and all $F_{ij}$ are independent.  Thus a deterministic mechanism $M=(x,p)$ satisfies
\begin{align} x_{ij}(t)\in \{0,1\}  \text{\ \ and \ } \sum_{\ell\in[m]}x_{i\ell}(t)\leq 1,\  \sum_{\ell\in[n]}x_{\ell j}(t)\leq 1  \text{\ \ for all \ } i, j.
\end{align}
Let $DREV^{(UD)}(\mathcal{F})$ be the maximum revenue achievable under $\mathcal{F}$ by any deterministic DIC-IR mechanism. Let $REV^{(UD)}(\mathcal{F})$ denote the maximum revenue achievable under $\mathcal{F}$ by any randomized BIC-BIR mechanism (equivalent to incentive-compatible lottery-based mechanisms \cite{CMS2015}).

An interesting connection was established in Chawla et al. \cite{CHMS2010}\cite{CMS2015} between unit-demand auctions and single-parameter environment auctions.  A unit-demand auction with valuation distribution $\mathcal{F}$ induces a single-parameter auction $OPT^{COPIES-UD}$ for $nm$ buyers $B_{ij}$ with independent valuation distributions $F_{ij}$, satisfying the same allocation constraints as Eq. 1.  Let $OPT^{COPIES-UD}(\mathcal{F})$ denote the maximum revenue achievable under $\mathcal{F}$.

\vskip 8pt
\noindent{\textbf{Theorem A}} (Chawla et al.\cite{CHMS2010}, Kleinberg and Weinberg \cite{KW2012})
\begin{align*}DREV^{UD}(\mathcal{F}) \leq OPT^{COPIES-UD}(\mathcal{F}) \leq 6\cdot DREV^{UD}(\mathcal{F}).
\end{align*}

The upper bound $ 6\cdot DREV^{UD}(\mathcal{F})$ in Theorem A is accomplished by certain sequential posted-price mechanism.

\vskip 8pt
\noindent{\textbf{Theorem B}} (Chawla, Malec and Sivan \cite{CMS2015}, Cai, Devanur and Weinberg \cite{CDWSTOC2016})
\begin{align*}REV^{UD}(\mathcal{F}) \leq 4\cdot OPT^{COPIES-UD}(\mathcal{F}).
\end{align*}

\noindent{\emph{Proof of Theorem 2}.}  From Theorem A and B, we have
\begin{align*}
DREV^{UD}(\mathcal{G})&\geq \frac{1}{6} OPT^{COPIES-UD}(\mathcal{G}),\\
OPT^{COPIES-UD}(\mathcal{F}) &\geq \max\{DREV^{UD}(\mathcal{F}), \frac{1}{4} REV^{UD}(\mathcal{F})\}
\end{align*}
Now, $OPT^{COPIES-UD}$ is a single-parameter environment auction , and thus by Theorem 1,
\begin{align*}
OPT^{COPIES-UD}(\mathcal{G}) \geq OPT^{COPIES-UD}(\mathcal{F}).
\end{align*}
Theorem 2 follows from all the above inequalities.  \qed

\section{Subadditive Combinatorial Auctions}

\subsection{Background}

We consider revenue maximization in the combinatorial auction with $n$ independent buyers and $m$ heterogeneous items.  The auction is specified by a \emph{type profile distribution}  $\mathcal{F}$ and \emph{valuation} function $v$ as elaborated below. We follow the convention used in \cite{CZ2017}\cite{RW}.

For each $i\in [n]$, buyer $i$ receives a \emph{type} $t_i=(t_{i1}, \cdots, t_{im})$, where $t_i$ is drawn from a product distribution
$F_i=F_{i1}\times \cdots \times F_{im}$ over some \emph{type space} $T_i=T_{i1}\times \cdots \times T_{im}$.  Let $\mathcal{F}=F_{1}\times \cdots \times F_{n}$ and $t=(t_{1}, \cdots, t_{m})$.  Let $T=T_{1}\times \cdots \times T_{n}$.  We also regard the type profile $t\in T$ as as an $n\times m$ matrix $t=(t_{ij})$, and $\mathcal{F}=(F_{ij})$ as an $n\times m$ matrix of independent distributions. For each $i\in [n]$, buyer $i$ has a \emph{valuation} function $v_i: T_i\times 2^{[m]}\rightarrow[0, \infty)$.  The quantity $v_i(t_i, S)$ expresses the value buyer $i$ attaches to the collection of items $S\subseteq [m]$ if $t_i$ is buyer $i$'s received type.  Let $v=(v_{1}, \cdots, v_{n})$.

A \emph{deterministic} mechanism $M=(x, p)$ specifies the \emph{allocation} $x=(x_{1}, \cdots, x_{n})$ and \emph{payment} $p=(p_{1}, \cdots, p_{n})$, where $x_i: T\rightarrow 2^{[m]}$ and $p_i: T\rightarrow [0, \infty)$, satisfying the condition that no item can be allocated to more than $1$ buyer. A \emph{randomized} mechanism is specified by a distribution over the set of deterministic mechanisms. Let $REV_{DIC}(v, \mathcal{F})$ and $REV_{BIC}(v, \mathcal{F})$ be the maximum achievable revenue of any deterministic DIC-IR mechanism, and any randomized BIC-BIR mechanism, respectively.

We are interested in valuation function $v_i$ that have the following properties:

\noindent(1) \emph{No Extenalities:} For each $t_i\in T_i$ and $S\subseteq [m]$,  $v_i(t_i, S)$ depends only on $(t_{ij}|\, j\in S)$.

\noindent(2) \emph{Monotone:} For each $t_i\in T_i$ and $U\subseteq V\subseteq [m]$, $v_i(t_i, U)\leq v_i(t_i, V)$.

\noindent(3) \emph{Subadditive:} For each $t_i\in T_i$ and $U\subseteq V\subseteq [m]$, $v_i(t_i, U\cup V)\leq v_i(t_i, U) + v_i(t_i, V)$.

We are particularly interested in valuations $v_i$ that are \emph{XOS} (or called \emph{fractionally subadditive}).  Namely, $v_i(t_i, S)=\max_{k\in[K]}v_i^{(k)}(t_i, S)$ where $K$ is finite for each $k$ and $v_i^{(k)}(t_i, \cdot)$ is an \emph{additive} function, i.e., $v_i^{(k)}(t_i, S)=\sum_{j\in S}v_i^{(k)}(t_i, \{j\})$ for all $S\subseteq [m]$.

In Rubinstein and Weinberg \cite{RW}, it was shown that, in the case of $1$ buyer ($n=1$), for any subadditive valuation and $\mathcal{F}$, there is a simple mechanism achieving a constant approximation of optimal revenue. It was also shown in Cai and Zhao \cite{CZ2017} that for any $n$, $\mathcal{F}$, and XOS valuation $v$, there exist certain sequential posted price mechanisms that achieve a constant approximation of optimal revenue. We review below some fact from \cite{CZ2017} that are essential for our proof of Theorem 3.

\vskip 8pt
\noindent{\textbf{Definition }}\  \emph{Rational Sequential Posted Price Mechanism (RSPM)}\\
Let $\xi=(\xi_{ij})$ where $\xi_{ij}\geq 0$ is a posted price for buyer $i$ and item $j$. Let $p=(p_{ij})$ where $p_{ij}\geq 0$ is payment of buyer $i$ if item $j$ is chosen. For $i=1$ to $n$, buyer $i$ picks at most $1$ item $j$ from the set $S$ of available items, maximizing buyer $i$'s utility $V_{ij}-\xi_{ij}$ (where $V_{ij}=v_i(t_i, \{j\}))$ is a function of $t_{ij}$); buyer $i$ pays $p_{ij}$ if $j$ is picked, and pays $0$ if no item is picked. We use $RSPM_{\xi}$ to denote this mechanism.

\vskip 8pt
\noindent{\textbf{Definition}}\  \emph{Anonymous Sequential Posted Price with Entry Fee Mechanism (ASPE)}\\
This mechanism is specified by $Q=(Q_{1}, \cdots, Q_{m})$,   $\delta=(\delta_{1}, \cdots, \delta_{n})$, where for each $j\in[m]$, $Q_j$ is \emph{posted price} for item $j$, and for each $i\in[n]$, $\delta_i(t_i, S)$ is the \emph{entry fee} paid by buyer $i$ with reported type $t_i$ when the set of available items is $S\subseteq [m]$.  The mechanism proceeds as follows.  For $i=1$ to $n$, buyer $i$ either picks no item (and pays $0$), or picks a subset $I$ from the set $S\subseteq [m]$ of still available items and pays $\delta_i(t_i, S)+\sum_{j\in I} Q_j$; we update $S$.  The chosen subset $I$ is determined by maximizing buyer $i$'s utility $v(t_i, I)-(\delta_i(t_i, S)+\sum_{j\in I} Q_j)$ among all $I\subseteq S$; ties are broken arbitrarily.  We use $ASPE_{Q,\delta}$ to denote this mechanism.

\vskip 8pt
\noindent{\textbf{Definition}}\  \emph{Supporting Price} (Dobzinski, Nisan and Schapira \cite{DNS2005})\\
For any $\alpha \geq 1$, a type $t_i$ and a subset $S\subseteq [m]$, prices $(p_j|\, j\in S)$ are $\alpha$\emph{-supporting prices} for $v_i(t_i, S)$ if \\
(1) $v_i(t_i, S')\geq \sum_{j\in S'} p_j$ for all $S'\subseteq S$, and\\
(2) $\sum_{j\in S} p_j\geq   \frac{1}{\alpha}v_i(t_i, S)$.

\vskip 8pt
\noindent{\textbf{Definition}} Given any valuation $v$, let $\alpha_v$ be such that for all $i$, $t_i$ and $S$, there exist $\alpha_v${-supporting prices} for $v_i(t_i, S)$.  Clearly, for XOS valuations $v_i$, we can take $\alpha_v=1$.  It is also known that for any subadditive valuations $v$, we can take $\alpha_v= O(\log m)$.

\vskip 8pt
\noindent{\textbf{Theorem D}} (Cai and Zhao \cite{CZ2017}) For any $v, \mathcal{F}$ and constant $b\in (0,1)$, there exist $\xi, Q, \delta$ such that
\begin{align*} REV_{BIC}(v, \mathcal{F})&\leq (12+ \frac{8}{1-b}) RSPM_{\xi}(v, \mathcal{F}) + 8 \alpha_v\sum_{j\in [m]} Q_j,\\
   ASPE_{Q, \delta}(v, \mathcal{F})&\geq\frac{1}{4}\sum_{j\in [m]} Q_j - C\cdot RSPM_{\xi}(v, \mathcal{F}).
\end{align*}
where $C=\frac{5}{2(1-b)} + \frac {b+1}{2b(1-b)}$.

\subsection{Proof of Theorem 3}

To prove Theorem 3, our plan is to prove the following two lemmas for the  $\{\xi, Q, \delta\}$ satisfying Theorem D above.  As $REV_{DIC}(v,\mathcal{G})\geq ASPE_{Q, \delta}(v, \mathcal{G})$,  these lemmas together with Theorem D immediately imply Theorem 3.

\vskip 8pt
\noindent{\textbf{Lemma 1}}  If $\mathcal{G}\succeq_v \mathcal{F}$, then
 $RSPM_{\xi}(v, \mathcal{F}) \leq 6 REV_{DIC}(v, \mathcal{G})$.

\vskip 8pt
\noindent{\textbf{Lemma 2}}  If $\mathcal{G}\succeq_v \mathcal{F}$, then
$ASPE_{Q, \delta}(\mathcal{G}) \geq \frac{1}{4} \sum_{j\in [m]} Q_j - C\cdot RSPM_{\xi}(v, \mathcal{F})$.

\vskip 8pt
\noindent{\emph{Proof of Lemma 1.}}  First we observe that $RSPM_{\xi}$ (under $v, \mathcal{F}$) can be regarded as a mechanism for a standard unit-demand auction where buyer $i$ has valuation $y_{ij}\in [0, \infty)$ with distribution $Y_{ij}=\{y_{ij}=v_i(t_i, \{j\})\, | \, t_i\sim F_i\}$.  Let $\mathcal{F}_v$ denote the product distribution of all $Y_{ij}$.  It is easy to see that $RSPM_{\xi}(v, \mathcal{F}) \leq DREV^{(UD)}(\mathcal{F}_v)$.  Now, as $\mathcal{G}\succeq_v \mathcal{F}$, we have $\mathcal{G}_v\succeq \mathcal{F}_v$.  By Theorem 2, we have
\begin{align*}DREV^{(UD)}(\mathcal{F}_v)\leq 6 DREV^{(UD)}(\mathcal{G}_v)\leq 6 REV_{DIC}(v, \mathcal{G}).
\end{align*} Lemma 1 follows immediately. \qed

\vskip 8pt
The rest of Section 4 is denoted to the proof of Lemma 2. We first define some notations related to the operations of $ASPE_{Q, \delta}$.  For any $i\in [n]$ and type profile $t$, let $t_{<i}=(t_1, \ldots, t_{i-1})$ and denote by $S_i(t_{<i})$ the \emph{available set} of items (i.e., not purchased by buyers $1, \ldots, i-1$) for buyer $i$ to choose from. For any $t_i$ and $I\subseteq [m]$, let $u_i(t_i, I)=\max_{I'\subseteq I}(v(t_i, I')-\sum_{j\in I'} Q_j)$.  By definition of $ASPE_{Q, \delta}$, if $u_i(t_i, I)<\delta_i(I)$ where $I=S_i(t_{<i})$, then buyer $i$ receives no item and pays $0$; if $u_i(t_i, I)\geq \delta_i(I)$, buyer $i$ receives a bundle $I'\subseteq I$ maximizing $v(t_i, I')-\sum_{j\in I'} Q_j$ and pays $\delta_i(I) + \sum_{j\in I'} Q_j$ (regarded as \emph{entry fee} $\delta_i(I)$ plus \emph{item price} $\sum_{j\in I'} Q_j$).  Let $SOLD(t)$ denote the set of items not picked by any buyer after $ASPE_{Q, \delta}$ is finished.

For any distribution $\mathcal{G}$, the revenue $ASPE_{Q, \delta}(\mathcal{G})$ consists of two parts: $EntryFee(\mathcal{G})$ and $ItemPrice(\mathcal{G})$ which are the expected value of entry fees and item prices, respectively, paid by all buyers.

We are now ready to prove Lemma 2.   We show that
\begin{align}
EntryFee(\mathcal{G})&\geq \frac{1}{4} \sum_{j\in [m]}Pr_{t'\sim \mathcal{G}}\{j\not\in SOLD(t')\}\cdot Q_j - C\cdot RSPM_{\xi}(v, \mathcal{F})\\
ItemPrice(\mathcal{G})&\geq  \sum_{j\in [m]}Pr_{t'\sim \mathcal{G}}\{j\in SOLD(t')\}\cdot Q_j
\end{align}
Lemma 2 follows immediately from Eq. 2, 3 by adding these inequalities together.

Eq. 3 is obvious by definition.  We now prove Eq. 2.  First recall the following result from \cite{CZ2017}.

\vskip 8pt
\noindent{\textbf{Lemma CZ}} (Cai and Zhao \cite{CZ2017})
\begin{align*}
EntryFee(\mathcal{F})\geq &\frac{1}{4}\sum_{j\in [m]}Pr_{t\sim \mathcal{F}}\{j\not\in SOLD(t)\}\cdot Q_j\\
&-C\cdot RSPM_{\xi}(v, \mathcal{F}).
\end{align*}

Note that the righthand side of Eq. 2 and the righthand side of Lemma CZ have exactly the same form, although their numerical values may be quite different.  The key insight for proving Eq. 2 is to generalize Lemma CZ to a more relaxed setting: the \emph{Digital Goods}, where the type profile $t$'s distribution can be de-coupled from the distribution of the available sets $\{S_i(t_{<i})\}$.  We show that in this setting, it becomes possible to compare the two entry fees $EntryFee(\mathcal{F})$ and $EntryFee(\mathcal{G})$. We emphasize that we are not casting our original research problem in a new setting. Rather, we merely embed $\mathcal{F}$ and $\mathcal{G}$ in this larger space where we will be able to find a connecting path between them for the purpose of comparing their revenues under $ASPE_{Q,\delta}$. (For more information on Digital Goods, see e.g. Hartline and Karlin \cite{HK2007}.)

In what follows, $v$ and $\mathcal{F}$ are fixed, while $Q, \delta,  \xi$ are determined by $v$, $\mathcal{F}$ (as specified in \cite{CZ2017}).

\vskip 8pt
\noindent{\textbf{Digital Goods (DG):}}

\noindent In this setting, each item $j\in [m]$ has an unlimited supply of identical copies, so that $j$ may be assigned to many buyers if necessary.  Let $\mathcal{I}=(\mathcal{I}_1, \ldots, \mathcal{I}_n)$  where each $\mathcal{I}_i$ is a distribution over $2^{[m]}$.  Consider the following mechanism: given a type profile $t=(t_1, \ldots, t_n)$, the seller generates for each buyer $i\in [n]$ a random $I_i\sim \mathcal{I}_i$ and applies $ASPE_{Q,\delta}$ to buyer $i$ with $I_i$ as the available set of items.  Thus, the buyer pays an entry fee $\delta_i(I_i)$ (and the appropriate item prices) if the condition $u_i(t_i, I_i)\geq \delta_i(I_i)$ is satisfied; otherwise buyer $i$ pays nothing and receives no items.  For any $\mathcal{I}$ and type profile distribution $\mathcal{H}$, let $EntryFee(\mathcal{H}, \mathcal{I})$ denote the expected total entry fees collected when $t\sim \mathcal{H}$, that is,
\begin{align*}
EntryFee(\mathcal{H}, \mathcal{I})= \sum_{i\in [n]}E_{t_i\sim H_i}\left[Pr_{I_i\sim \mathcal{I}_i}\{u_i(t_i, I_i)\geq\delta_i(I_i)\}\cdot \delta_i(I_i)\right].
\end{align*}
\noindent{\textbf{Definition} \  [Embedding of $\mathcal{H}$ in DG Space] \
For any type profile distribution $\mathcal{H}$, let $\mathcal{I}^{\mathcal{H}}=(\mathcal{I}^{\mathcal{H}}_1, \ldots, \mathcal{I}^{\mathcal{H}}_n)$  where each $\mathcal{I}^{\mathcal{H}}_i$  is $\{S_i(t_{<i})\,|\, t\sim \mathcal{H}\}$.
It is obvious that $\mathcal{H} \rightarrow (\mathcal{H}, \mathcal{I}^{\mathcal{H}})$ is an embedding that preserves Entryfee, that is,
\begin{align*}EntryFee(\mathcal{H})=EntryFee(\mathcal{H}, \mathcal{I}^{\mathcal{H}}).
\end{align*}

We state below the embeddings of our targeted distributions $\mathcal{F}$,  $\mathcal{G}$ for easy reference.
\vskip 8pt
\noindent{\textbf{Fact 1}} \ \ $EntryFee(\mathcal{F})=EntryFee(\mathcal{F}, \mathcal{I}^{\mathcal{F}}) \ \ \text{where\ \  }  \mathcal{I}^{\mathcal{F}}_i \text{ \ is  \ } \{S_i(t_{<i})\,|\, t\sim \mathcal{F}\}$, \\
$\hbox{}\quad\quad\quad\quad\  EntryFee(\mathcal{G})=EntryFee(\mathcal{G}, \mathcal{I}^{\mathcal{G}})\ \ \ \text{where\ \ }  \mathcal{I}^{\mathcal{G}}_i \text{\ \ is  \ } \{S_i(t'_{<i})\,|\, t'\sim \mathcal{G}\}$.

Fact 1 suggests that,  $EntryFee(\mathcal{F})$ and $EntryFee(\mathcal{G})$ may be compared in the DG space via a connecting point between $(\mathcal{F}, \mathcal{I}^{\mathcal{F}})$ and $(\mathcal{G}, \mathcal{I}^{\mathcal{G}})$, such as $(\mathcal{F}, \mathcal{I}^{\mathcal{G}})$. This turns out to be indeed the case as seen below.

\vskip 8pt
\noindent{\textbf{Lemma 3}} [Monotonicity]   \ For any $\mathcal{I}$ and $\mathcal{G} \succeq_v \mathcal{F}$,
\begin{align*}EntryFee(\mathcal{G}, \mathcal{I})\geq EntryFee(\mathcal{F}, \mathcal{I}).
\end{align*}

\noindent{\textbf{Lemma 4}} [Digital Goods Extension of CZ Lemma] \  For any $\mathcal{I}$,
\begin{align*}EntryFee(\mathcal{F}, \mathcal{I})\geq  \frac{1}{4} \sum_{j\in [m]}B_j\cdot Q_j - C\cdot RSPM_{\xi}(v, \mathcal{F})
\end{align*}
where $B_j=\min_{i\in [n]}\left[Pr_{I_i\sim \mathcal{I}_i}\{j\in I_i\}\right]$.

\vskip 8pt
\noindent{\emph{Proof of Lemma 3.}} Write $t'_i \geq_v t_i$ if $v(t'_i, S)\geq  v(t_i, S)$ for all $S\subseteq [m]$.  By definition of $v$-domination, one can generate a random pair of types $(t'_i, t_i)$ such that (a) $t'_i \geq_v t_i$, and (b) marginal distribution of $t'_i, t_i$ equals $G_i, F_i$ respectively.  Noting that $u_i(t'_i, I_i)\geq u_i(t_i, I_i)$ whenever $t'_i \geq_v t_i$, we have
\begin{align*}EntryFee(\mathcal{G}, \mathcal{I})&=\sum_{i\in [n]}E_{t'_i\sim G_i}\left[Pr_{I_i\sim \mathcal{I}_i}\{u_i(t'_i, I_i)\geq\delta_i(I_i)\}\cdot \delta_i(I_i)\right]\\
&\geq\sum_{i\in [n]}E_{t_i\sim F_i}\left[Pr_{I_i\sim \mathcal{I}_i}\{u_i(t_i, I_i)\geq\delta_i(I_i)\}\cdot \delta_i(I_i)\right]\\
&=EntryFee(\mathcal{F}, \mathcal{I})
\end{align*}\qed

\noindent{\emph{Proof of Lemma 4.}} This requires a lengthy and complex proof.  Fortunately, the proof of Lemma CZ as given in \cite{CZ2017} (in the arXiv full paper version, Lemma 31-34) can be extended line-by-line to our Digital Goods Setting with minimum (and obvious) modifications to yield the present lemma, and hence will not be repeated here.  \qed

\vskip 8pt
\noindent{\textbf{Fact 2}} \ For $\mathcal{I}_i=\{S_i(t'_{<i})\, |\,t'\sim \mathcal{G}\}$, we have
\begin{align*}B_j\geq Pr_{t'\sim \mathcal{G}}\{j\not\in SOLD(t')\}.
\end{align*}

\noindent{\emph{Proof.}} \ Obvious.  \qed

\vskip 8pt
It follows from Fact 1 and Lemma 3 that
\begin{align*}
EntryFee(\mathcal{G})\geq EntryFee(\mathcal{F}, \mathcal{I})
\end{align*}
where $\mathcal{I}_i=\{S_i(t'_{<i})\, |\,t'\sim \mathcal{G}\}$.

By Fact 2 and Lemma 4, we have then
\begin{align*}
EntryFee(\mathcal{G})\geq &\frac{1}{4} \sum_{j\in [m]}Pr_{t'\sim \mathcal{G}}\{j\not\in SOLD(t')\}\cdot Q_j \\
                          &- C\cdot RSPM_{\xi}(v, \mathcal{F}).
\end{align*}
This proves Eq. 2 and thus Lemma 2.  We have completed the proof of Theorem 3.

\end{document}